\documentclass[runningheads]{llncs}

\usepackage{eccv}

\usepackage{eccvabbrv}

\usepackage{graphicx}
\usepackage{booktabs}
\usepackage{multirow}

\usepackage{algorithm}
\usepackage{algorithmic}

\usepackage{caption}
\usepackage{wrapfig,lipsum}
\usepackage[accsupp]{axessibility}  % Improves PDF readability for those with disabilities.

\usepackage[pagebackref,breaklinks,colorlinks,citecolor=eccvblue]{hyperref}
\usepackage{orcidlink}

\begin{document}

\title{Do Not Leave a Gap: Hallucination-Free Object Concealment in Vision-Language Models} 

\titlerunning{Do Not Leave a Gap: Hallucination-Free Object Concealment in Vision-Language Models}

\author{Amira Guesmi \and
Muhammad Shafique} %\and
%Third Author\inst{3}\orcidlink{2222--3333-4444-5555}}

\authorrunning{A.~Guesmi et al.}

\institute{Engineering Division, New York University Abu Dhabi, UAE} %\and
% Springer Heidelberg, Tiergartenstr.~17, 69121 Heidelberg, Germany

\maketitle

\begin{abstract}

Vision-language models (VLMs) have recently shown remarkable capabilities in visual understanding and generation, but remain vulnerable to adversarial manipulations of visual content.
Prior object-hiding attacks primarily rely on suppressing or blocking region-specific representations, often creating semantic gaps that inadvertently induce hallucination, where models invent plausible but incorrect objects.
In this work, we demonstrate that hallucination arises not from object absence per se, but from semantic discontinuity introduced by such suppression-based attacks.
We propose a new class of \emph{background-consistent object concealment} attacks, which hide target objects by re-encoding their visual representations to be statistically and semantically consistent with surrounding background regions.
Crucially, our approach preserves token structure and attention flow, avoiding representational voids that trigger hallucination.
We present a pixel-level optimization framework that enforces background-consistent re-encoding across multiple transformer layers while preserving global scene semantics.
Extensive experiments on state-of-the-art vision-language models show that our method effectively conceals target objects while preserving up to $86\%$ of non-target objects and reducing grounded hallucination by up to $3\times$ compared to attention-suppression-based attacks.
Qualitative results further confirm that our approach maintains scene coherence and avoids spurious object insertion.
Our findings highlight semantic continuity as a key factor in hallucination behavior and introduce a new direction for adversarial analysis of generative multimodal models.
\keywords{VLM \and Hallucination \and Object concealment \and Adversarial attacks}
\end{abstract}
\section{Introduction}

Vision--Language Models (VLMs) have become foundational components in modern AI systems, enabling image captioning, visual question answering, and multimodal reasoning~\cite{radford2021clip, alayrac2022flamingo, li2023blip2}.
As these models are increasingly deployed in sensitive settings, there is growing interest in \emph{object concealment} and \emph{visual privacy} attacks, where specific regions or objects in an image are intentionally hidden from the model’s perception while preserving the overall semantic integrity of the scene~\cite{meftah2025vip, luo2024image, hu2024firm}. A dominant class of existing object concealment attacks operates by \emph{suppressing} visual information corresponding to a target region, for example by masking pixels, erasing patches, or explicitly reducing attention weights associated with region-of-interest (ROI) tokens~\cite{dai2023instructblip, zhang2024visual, meftah2025vip}.
While such approaches can successfully prevent the direct recognition of the target object, they frequently induce a secondary and often overlooked failure mode: \emph{hallucination}.
When queried about the scene, the model compensates for missing visual evidence by inventing objects, attributes, or relationships that were never present in the original image~\cite{liu2024survey, datta2025evaluating}.

\begin{figure}
    \centering
    \includegraphics[width=\linewidth]{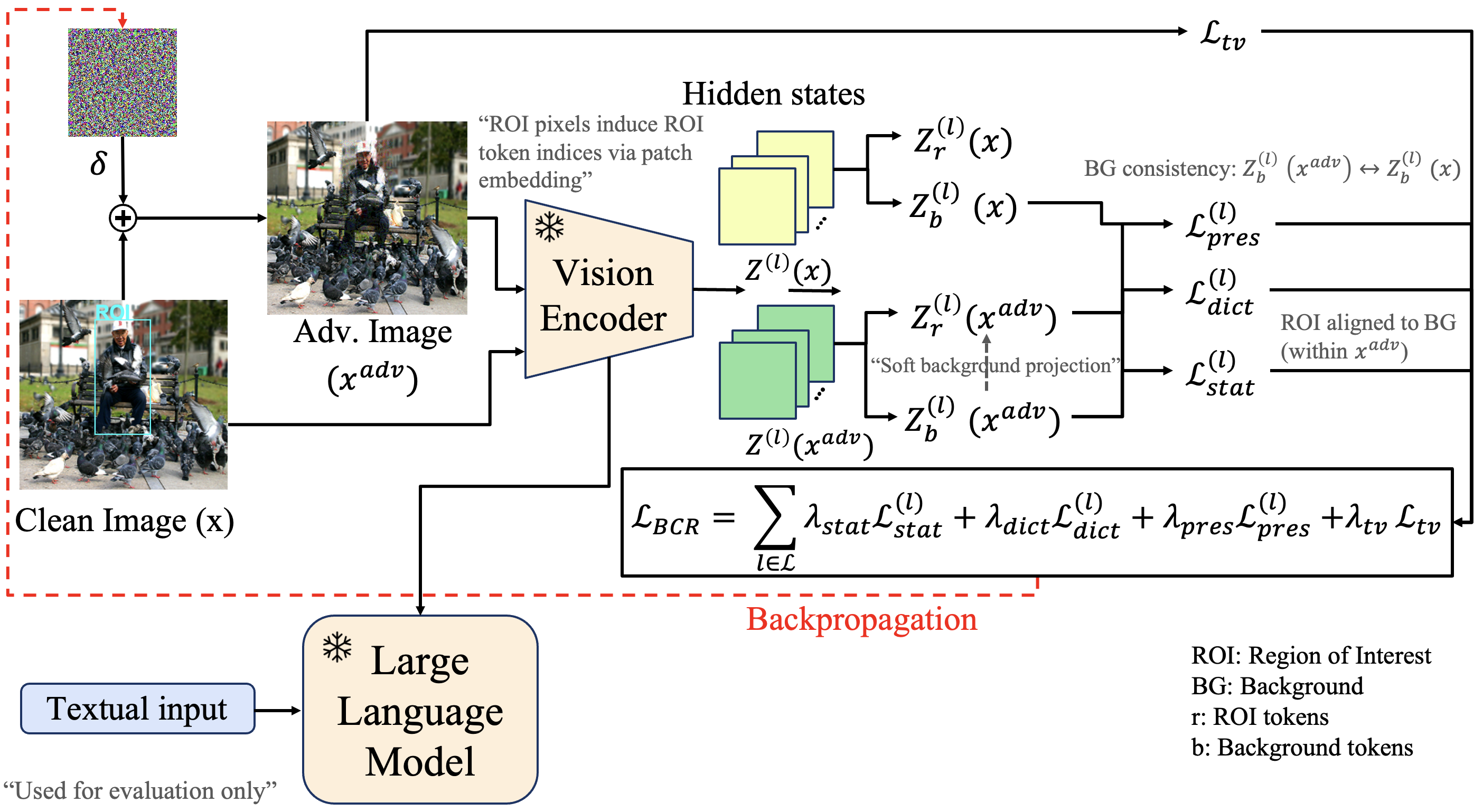}
    \caption{Overview of \textbf{Background-Consistent Re-encoding (BCR)}.
A pixel-level perturbation $\delta$ is optimized to produce an adversarial image $\mathbf{x}^{adv}$ from a clean image $\mathbf{x}$ with a specified ROI.
A frozen vision encoder extracts layer-wise hidden states, from which ROI and background tokens are identified via patch embedding.
BCR enforces semantic continuity by (i) aligning ROI and background statistics, (ii) softly projecting ROI features onto background representations, and (iii) preserving background tokens between clean and adversarial images.
A total variation regularizer encourages smooth perturbations.
The objective is optimized across multiple transformer layers, while the language model remains frozen and is used only for evaluation.
}
    \label{fig:method}
\end{figure}

This phenomenon is commonly treated as an unavoidable side effect of concealment.
In contrast, we argue that hallucination is not an incidental artifact of object removal, but a direct consequence of \emph{semantic discontinuity} introduced by suppression-based attacks.
By aggressively attenuating or nullifying ROI representations, prior methods create representational gaps within the vision encoder.
These gaps propagate through cross-modal alignment layers, prompting the language model to infer missing semantics and fill them with plausible—but incorrect—content~\cite{liu2024survey}.
In this work, we propose a different perspective: effective object concealment should preserve the \emph{structural and statistical consistency} of visual representations, even when the target object is hidden.
Instead of removing ROI tokens or severing their influence, we aim to \emph{re-encode} them such that they remain present but become indistinguishable from background content.
Under this formulation, the global image representation remains coherent, reducing the incentive for the language model to hallucinate missing entities~\cite{li2023blip2, dai2023instructblip}.

To operationalize this idea, we introduce \emph{Background-Consistent Re-encoding (BCR)} (see Figure \ref{fig:method}), a pixel-level attack framework that conceals objects by aligning the internal representations of ROI tokens with background statistics across multiple layers of the vision transformer.
BCR jointly enforces statistical similarity, dictionary-based projection onto background features, and preservation of non-ROI content, resulting in adversarial images that hide target objects without introducing semantic gaps.
Importantly, our approach operates directly in pixel space, requiring no modification of model parameters and remaining applicable to modern VLM architectures~\cite{goodfellow2014explaining, madry2017towards}.

% We evaluate BCR on state-of-the-art VLMs using object-aware and hallucination-sensitive metrics. Our results demonstrate that BCR achieves high concealment success while significantly reducing hallucination compared to suppression-based baselines, thereby revealing a previously underexplored trade-off between object hiding and semantic consistency in multimodal models.

% \paragraph{Contributions.}
This paper makes the following contributions:
\begin{itemize}
    % \item We identify semantic discontinuity induced by suppression-based attacks as a primary cause of hallucination in vision--language models.
    \item We introduce \emph{Background-Consistent Re-encoding (BCR)}, a novel object concealment paradigm that preserves token structure and attention flow while hiding target objects.
    \item We propose a principled pixel-level optimization framework that enforces background-consistent visual representations across multiple vision transformer layers.
    \item We design hallucination-aware evaluation metrics and empirically demonstrate that BCR substantially reduces hallucination while maintaining strong concealment performance across multiple VLM architectures.
\end{itemize}

\section{Related Work}

\subsection{Adversarial Attacks on Vision--Language Models}

Vision--language models (VLMs) have been shown to inherit and amplify adversarial vulnerabilities from their visual backbones while introducing new cross-modal failure modes. Early work adapted classical image-space attacks such as FGSM and PGD~\cite{goodfellow2014explaining,madry2017towards} to multimodal architectures, demonstrating that small, imperceptible perturbations can significantly degrade image captioning and visual question answering (VQA) performance~\cite{zhang2024visual,luo2024image}. These studies established that multimodal grounding does not inherently confer robustness and that adversarial perturbations can propagate through cross-modal alignment layers to corrupt downstream language generation.
Subsequent research has explored more structured attack strategies that explicitly exploit multimodal fusion mechanisms. %Prompt-transfer attacks reveal that adversarial perturbations crafted under one textual query often generalize across diverse prompts, exposing vulnerabilities in shared visual representations~\cite{luo2024image}. 
While effective, these attacks generally aim to \emph{break} the model’s predictions, often at the cost of destroying overall image semantics and utility.
%Representation-level attacks manipulate hierarchical patch embeddings or intermediate features to induce global semantic corruption without directly targeting output tokens~\cite{hu2024firm}. Other approaches modify attention distributions or alignment modules to suppress region-specific evidence, effectively preventing object recognition while often introducing semantic discontinuities~\cite{meftah2025vip}. 

% While these attacks are effective at degrading model performance or suppressing sensitive content, they primarily aim to \emph{break} predictions by disrupting visual evidence or cross-modal alignment. 
% This frequently leads to collateral semantic degradation, unstable language generation, or hallucinated content. In contrast, our work departs from suppression-based paradigms and instead investigates continuity-preserving adversarial manipulation, where target objects are concealed through background-consistent re-encoding rather than representational voids.

\subsection{Adversarial Attacks for Privacy and Information Protection}
More recently, adversarial attacks have been explored as tools for privacy preservation.
VIP (Visual Information Protection) \cite{meftah2025vip} frames privacy as an adversarial objective and proposes selectively masking regions of interest (ROIs) by suppressing attention and value activations in early vision layers, effectively preventing VLMs from recognizing sensitive content. 
VIP demonstrates strong concealment performance across multiple VLMs, but relies on explicit ROI suppression, which creates a representational ``gap'' that can encourage the language model to hallucinate or infer missing content. Other methods explicitly manipulate internal representations rather than output tokens. For instance,
PRM (Patch Representation Misalignment) \cite{hu2024firm} disrupts hierarchical patch representations by enforcing feature divergence between clean and adversarial images, leading to global semantic corruption.
However, these methods typically alter pixel-level appearance or remove information entirely, making them unsuitable for scenarios where global semantic coherence must be preserved.

% \subsection{Contrast with Our Approach}
In contrast to prior work, our method does not suppress or erase visual information within the ROI.
Instead, we propose \emph{Background-Consistent Re-encoding (BCR)} manipulation, which reshapes the \emph{distributional relationship} between ROI and background tokens inside the vision encoder.
% By enforcing statistical similarity, dictionary overlap constraints, and background preservation across multiple layers, BCR conceals target objects \emph{without creating explicit representational gaps}.
% This prevents downstream language models from hallucinating missing objects while preserving the overall semantic structure of the image.

% To the best of our knowledge, this is the first work to explicitly study object concealment in VLMs through representation blending rather than suppression, offering a new perspective on privacy-preserving adversarial attacks that maintain semantic continuity.

\section{Threat Model}
We consider a \emph{white-box, image-level adversary} targeting vision--language models (VLMs).
The adversary is given access to the model architecture and parameters, and can compute gradients with respect to the visual input.
The attack operates by applying a bounded perturbation to the input image at the pixel level, subject to an $\ell_\infty$ norm constraint.
The adversary is additionally provided with a region of interest (ROI), specified as a bounding box, corresponding to an object whose visual presence should be concealed.
Such ROIs naturally arise from object detectors, annotations, or user-defined sensitive regions.

The adversary's objective is not to induce misclassification or nonsensical outputs, but rather to \emph{remove the semantic evidence of the target object} from the model’s internal representation while preserving global visual coherence.
Importantly, the attacker seeks to avoid introducing explicit visual artifacts or representational gaps that could trigger compensatory hallucinations by the language decoder.
Unlike object removal or inpainting approaches, our objective is not to alter the visual scene but to modulate representational grounding within the vision–language model. This enables privacy-preserving concealment while maintaining perceptual readability and evidentiary integrity for human observers.
The adversary does not modify the model weights, prompts, or decoding strategy, and has no control over the language model beyond its dependence on visual features.
All evaluations are performed under this fixed threat model.

\section{Attack Principle}
\label{sec:attack_principle}

% \paragraph{Problem Setting.}
% Vision--language models (VLMs) rely on a continuous visual representation to ground language generation.
% When salient visual evidence is abruptly removed or heavily suppressed (e.g., by masking, erasing, or zeroing a localized region) the model is forced to infer missing content from contextual priors.
% This inference process frequently manifests as \emph{hallucination}, where the model fabricates objects or attributes that are not present in the image.
% Existing object-hiding attacks often rely on explicit removal or severe suppression of visual features corresponding to a target region.
% While effective at concealing the object itself, such approaches introduce a representational ``gap'' that the model attempts to compensate for during decoding, leading to unstable or hallucinated outputs.
% This effect is particularly pronounced in large VLMs, where strong language priors amplify missing visual evidence.
\noindent\textit{Problem Setting.}
Vision--language models (VLMs) rely on a continuous visual representation to ground language generation, where aligned visual tokens provide the evidential basis for downstream captioning and reasoning~\cite{radford2021clip,alayrac2022flamingo,li2023blip2}. 
When salient visual evidence is abruptly removed or heavily suppressed (e.g., via masking, erasing, or zeroing localized regions), the model is forced to infer missing content from contextual priors encoded in the language decoder~\cite{rohrbach2018object,liu2024survey}. 
This inference process frequently manifests as \emph{hallucination}, where the model fabricates objects or attributes that are not visually present. 
Existing object-hiding or privacy-oriented attacks often rely on explicit removal or severe suppression of visual features corresponding to a target region~\cite{meftah2025vip,zhang2024visual}. While effective at concealing the object itself, such approaches introduce a representational ``gap'' in the visual embedding space that the model attempts to compensate for during decoding, leading to unstable or hallucinated outputs~\cite{liu2024survey}. This effect is particularly pronounced in large-scale VLMs, where strong language priors and generative biases amplify missing or ambiguous visual evidence~\cite{liu2024survey}.

% \paragraph{Key Insight.}
% We observe that hallucination is not caused by concealment per se, but by \emph{representation discontinuity}.
% Specifically, when the feature distribution of a target region deviates sharply from the surrounding background, the global image representation (e.g., the CLS token or pooled visual embedding) encodes this discrepancy as an anomaly.
% During generation, the language model resolves this anomaly by hypothesizing plausible but incorrect objects.

\noindent\textit{Principle of Background-Consistent Re-encoding (BCR).}
Instead of removing or suppressing the target object, we propose to \emph{re-encode} its visual representation such that it becomes statistically and semantically consistent with the surrounding background.
The goal is not to erase information, but to \emph{blend} the target region into the background feature manifold.
Concretely, BCR enforces three complementary constraints: \\
% \begin{itemize}
%     \item \textbf{Statistical alignment}: The first- and second-order feature statistics of target-region tokens are matched to those of background tokens.
%     \item \textbf{Dictionary consistency}: Target-region features are reconstructed as convex combinations of background features, ensuring that they lie within the background feature span.
%     \item \textbf{Background preservation}: Features outside the target region are explicitly preserved to prevent global degradation or collateral semantic drift.
% \end{itemize}
\noindent \textbf{- Statistical alignment}: The first- and second-order feature statistics of target-region tokens are matched to those of background tokens.\\
\noindent \textbf{- Dictionary consistency}: Target-region features are reconstructed as convex combinations of background features, ensuring that they lie within the background feature span. \\
\noindent \textbf{- Background preservation}: Features outside the target region are explicitly preserved to prevent global degradation or collateral semantic drift.

By maintaining continuity in the visual feature space, the global image representation remains coherent and does not signal missing content to the language decoder.
As a result, the model neither mentions the concealed object nor compensates for its absence via hallucination.

\noindent\textit{Contrast with Suppression-Based Attacks.}
Unlike masking- or suppression-based attacks (e.g., patch removal, attention blocking, or zeroing), BCR does not introduce visually or representationally empty regions.
Instead, the target object is \emph{absorbed} into the background representation.
This distinction is critical: while suppression creates a void that invites hallucination, background-consistent re-encoding preserves representational smoothness and stabilizes downstream language generation.
The resulting adversarial image conceals the target object while preserving global semantics and minimizing hallucinated content.
Importantly, BCR operates entirely at the representation level while being implemented via pixel-space optimization, making it applicable to black-box or frozen VLMs without modifying model parameters.

\section{Attack Formulation}
\label{sec:attack_formulation}

We consider a vision--language model (VLM) composed of a vision encoder $f_{\theta}$ and a language decoder $g_{\phi}$.
Given an input image $\mathbf{x} \in \mathbb{R}^{3 \times H \times W}$, the vision encoder produces a sequence of visual tokens
$\mathbf{Z} = f_{\theta}(\mathbf{x}) \in \mathbb{R}^{T \times D}$, where the first token corresponds to a global representation (CLS) and the remaining tokens encode spatial image regions.
The language decoder conditions on this representation to generate a caption or answer.
Let $\mathcal{R}$ denote a target region of interest (ROI) corresponding to an object that the attacker wishes to conceal, specified by bounding boxes.
Let $\mathcal{I}_r \subset \{1,\dots,T-1\}$ be the indices of visual tokens whose receptive fields intersect the ROI, and $\mathcal{I}_b$ the indices of background tokens.
Our goal is to produce an adversarial image $\mathbf{x}^{\mathrm{adv}}$ that conceals the target object while preserving global semantic consistency and avoiding hallucination.

\subsection{Background-Consistent Re-encoding Objective}

Unlike suppression-based attacks that remove or mask ROI features, we aim to \emph{re-encode} ROI tokens such that they become statistically and semantically indistinguishable from background tokens.
Formally, we optimize the following objective:
\begin{equation}
\label{eq:bcr_objective}
\min_{\mathbf{x}^{\mathrm{adv}}} 
\;\; 
\mathcal{L}_{\mathrm{BCR}}(\mathbf{x}^{\mathrm{adv}})
=
\sum_{l \in \mathcal{L}}
\Big(
\lambda_{\mathrm{stat}} \mathcal{L}_{\mathrm{stat}}^{(l)}
+
\lambda_{\mathrm{dict}} \mathcal{L}_{\mathrm{dict}}^{(l)}
+
\lambda_{\mathrm{pres}} \mathcal{L}_{\mathrm{pres}}^{(l)}
\Big)
+
\lambda_{\mathrm{tv}} \mathcal{L}_{\mathrm{tv}},
\end{equation}
% \begin{equation}
% \label{eq:bcr_objective}
% \begin{aligned}
% \min_{\mathbf{x}^{\mathrm{adv}}} \;\;
% \mathcal{L}_{\mathrm{BCR}}(\mathbf{x}^{\mathrm{adv}})
% =
% &\sum_{l \in \mathcal{L}}
% \Big(
% \lambda_{\mathrm{stat}} \mathcal{L}_{\mathrm{stat}}^{(l)}
% +
% \lambda_{\mathrm{dict}} \mathcal{L}_{\mathrm{dict}}^{(l)}
% + \\
% &\quad\lambda_{\mathrm{pres}} \mathcal{L}_{\mathrm{pres}}^{(l)}
% \Big)
%  + \lambda_{\mathrm{tv}} \mathcal{L}_{\mathrm{tv}} .
% \end{aligned}
% \end{equation}

subject to a pixel-level perturbation budget
$\|\mathbf{x}^{\mathrm{adv}} - \mathbf{x}\|_{\infty} \leq \epsilon$.
The losses are computed across a set of intermediate vision encoder layers $\mathcal{L}$ to enforce multi-level consistency.

\subsection{Statistical Alignment Loss}

To eliminate distributional discrepancies that could signal the presence of a concealed object, we align the first- and second-order statistics of ROI and background features.
For layer $l$, let $\mathbf{Z}_r^{(l)} = \{\mathbf{z}_i^{(l)} \mid i \in \mathcal{I}_r\}$ and
$\mathbf{Z}_b^{(l)} = \{\mathbf{z}_j^{(l)} \mid j \in \mathcal{I}_b\}$.
We define:
\begin{equation}
\label{eq:stat_loss}
\mathcal{L}_{\mathrm{stat}}^{(l)}
=
\big\|
\mu(\mathbf{Z}_r^{(l)}) - \mu(\mathbf{Z}_b^{(l)})
\big\|_2^2
+
\big\|
\sigma(\mathbf{Z}_r^{(l)}) - \sigma(\mathbf{Z}_b^{(l)})
\big\|_2^2,
\end{equation}
where $\mu(\cdot)$ and $\sigma(\cdot)$ denote per-dimension mean and standard deviation.
This loss enforces that ROI features follow the same distribution as background features, preventing detectable anomalies.

% \subsection{Dictionary Projection Loss}

% While statistical alignment controls low-order moments, it does not guarantee semantic consistency.
% To ensure ROI features lie within the background feature manifold, we introduce a dictionary projection loss.
% For each ROI token, we compute a soft assignment over background tokens using scaled dot-product similarity:
% \begin{equation}
% \alpha_{ij}^{(l)} =
% \frac{
% \exp\left( \mathbf{z}_{r,i}^{(l)} \cdot \mathbf{z}_{b,j}^{(l)} / \tau \right)
% }{
% \sum_{k \in \mathcal{I}_b}
% \exp\left( \mathbf{z}_{r,i}^{(l)} \cdot \mathbf{z}_{b,k}^{(l)} / \tau \right)
% },
% \end{equation}
% and project ROI features onto the background dictionary:
% \begin{equation}
% \hat{\mathbf{z}}_{r,i}^{(l)} = \sum_{j \in \mathcal{I}_b} \alpha_{ij}^{(l)} \mathbf{z}_{b,j}^{(l)}.
% \end{equation}
% The dictionary loss is then:
% \begin{equation}
% \label{eq:dict_loss}
% \mathcal{L}_{\mathrm{dict}}^{(l)}
% =
% \frac{1}{|\mathcal{I}_r|}
% \sum_{i \in \mathcal{I}_r}
% \big\|
% \mathbf{z}_{r,i}^{(l)} - \hat{\mathbf{z}}_{r,i}^{(l)}
% \big\|_2^2.
% \end{equation}
% This encourages ROI tokens to be expressed as convex combinations of background tokens, effectively re-encoding object features as background-consistent semantics.
\subsection{Dictionary Projection Loss}

While statistical alignment controls low-order moments, it does not ensure semantic consistency.
To explicitly constrain ROI features to lie on the background feature manifold, we introduce a soft background projection loss.
For each ROI token, we compute a soft assignment over background tokens using scaled dot-product similarity: $\alpha_{ij}^{(l)} =
\frac{
\exp\left( \mathbf{z}_{r,i}^{(l)} \cdot \mathbf{z}_{b,j}^{(l)} / \tau \right)
}{
\sum_{k \in \mathcal{I}_b}
\exp\left( \mathbf{z}_{r,i}^{(l)} \cdot \mathbf{z}_{b,k}^{(l)} / \tau \right)
}$. 
% \begin{equation}
% \alpha_{ij}^{(l)} =
% \frac{
% \exp\left( \mathbf{z}_{r,i}^{(l)} \cdot \mathbf{z}_{b,j}^{(l)} / \tau \right)
% }{
% \sum_{k \in \mathcal{I}_b}
% \exp\left( \mathbf{z}_{r,i}^{(l)} \cdot \mathbf{z}_{b,k}^{(l)} / \tau \right)
% }.
% \end{equation}
Each ROI feature is then projected onto the background feature dictionary: $\hat{\mathbf{z}}_{r,i}^{(l)} = \sum_{j \in \mathcal{I}_b} \alpha_{ij}^{(l)} \mathbf{z}_{b,j}^{(l)}.$
% \begin{equation}
% \hat{\mathbf{z}}_{r,i}^{(l)} = \sum_{j \in \mathcal{I}_b} \alpha_{ij}^{(l)} \mathbf{z}_{b,j}^{(l)}.
% \end{equation}

The dictionary loss is defined as:
\begin{equation}
\mathcal{L}_{\mathrm{dict}}^{(l)}
=
\frac{1}{|\mathcal{I}_r|}
\sum_{i \in \mathcal{I}_r}
\big\|
\mathbf{z}_{r,i}^{(l)} - \hat{\mathbf{z}}_{r,i}^{(l)}
\big\|_2^2.
\end{equation}

This loss encourages ROI tokens to be re-encoded as convex combinations of background features, enforcing semantic continuity without removing token structure.

\subsection{Background Preservation Loss}

To avoid global semantic drift and unintended distortions, we explicitly preserve background features by penalizing deviations from their clean counterparts:
\begin{equation}
\label{eq:preserve_loss}
\mathcal{L}_{\mathrm{pres}}^{(l)}
=
\frac{1}{|\mathcal{I}_b|}
\sum_{j \in \mathcal{I}_b}
\big\|
\mathbf{z}_{b,j}^{(l)}(\mathbf{x}^{\mathrm{adv}})
-
\mathbf{z}_{b,j}^{(l)}(\mathbf{x})
\big\|_2^2.
\end{equation}
This constraint ensures that only the ROI representation is altered, while the remainder of the image remains perceptually and semantically stable.

\subsection{Pixel-Level Regularization}

Finally, we apply total variation regularization to encourage spatially smooth perturbations within the ROI:
\begin{equation}
\label{eq:tv_loss}
\mathcal{L}_{\mathrm{tv}}
=
\sum_{c,h,w}
\big|
\mathbf{x}^{\mathrm{adv}}_{c,h,w+1} - \mathbf{x}^{\mathrm{adv}}_{c,h,w}
\big|
+
\big|
\mathbf{x}^{\mathrm{adv}}_{c,h+1,w} - \mathbf{x}^{\mathrm{adv}}_{c,h,w}
\big|.
\end{equation}

% \paragraph{Discussion.}
By jointly enforcing statistical alignment, dictionary-based semantic projection, and background preservation across multiple layers, BCR removes object-specific information without introducing representational gaps.
As a result, the global image representation remains coherent, preventing both object mention and compensatory hallucination during language generation.

\noindent\textit{Why Representation Continuity Prevents Hallucination.}
Vision--language models rely on the assumption that visual token representations form a continuous and semantically coherent manifold.
During generation, the language decoder implicitly performs inference over this manifold: when a region-specific signal is missing or anomalous, the decoder compensates by extrapolating from learned visual--semantic priors, often resulting in hallucinated content.
This behavior is particularly pronounced in suppression-based attacks, where masking, zeroing, or attention blocking creates a representational ``gap'' that violates the continuity assumptions learned during pretraining. Our attack avoids this failure mode by enforcing \emph{representation continuity} rather than suppression.
By re-encoding ROI tokens to match the statistical distribution and semantic span of background tokens, we ensure that the visual embedding remains locally smooth and globally consistent.
From the decoder’s perspective, the ROI does not appear as missing information, but as background-consistent evidence.
Consequently, there is no incentive for the model to hypothesize or hallucinate alternative objects to explain an apparent void.
Importantly, this continuity is enforced across multiple layers of the vision encoder, preventing object semantics from re-emerging through higher-level abstractions or cross-token interactions.
Thus, hallucination is reduced not by restricting generation, but by eliminating representational discontinuities that would otherwise trigger compensatory reasoning in the language model.

\section{Evaluation Metrics}
\label{sec:metrics}

Evaluating object concealment attacks on vision--language models requires more than measuring caption similarity or task accuracy.
A successful concealment attack must simultaneously (i) remove evidence of a target object, (ii) preserve non-target visual semantics, and (iii) avoid inducing hallucinated content.
To capture these distinct objectives, we introduce a set of complementary evaluation metrics combining caption analysis with grounded visual verification.

\noindent \textbf{Object Sets.}
Given a clean image $I$ and its adversarial counterpart $I'$, we generate captions
$c = f(I)$ and $c' = f(I')$ using the same prompting strategy.
From each caption, we extract a set of object-like tokens using a dependency-based noun phrase parser:
\[
\mathcal{O}(c), \quad \mathcal{O}(c').
\]
All object strings are normalized (lemmatized and lowercased) prior to comparison.

\noindent \textbf{Concealment Success.}
Concealment Success measures whether the target object is successfully removed from the adversarial caption.
Let $o^\ast$ denote the target object.
We define:
\begin{equation}
% \[
\text{ConcealmentSuccess} =
\mathbb{1}\big[ o^\ast \in \mathcal{O}(c) \;\land\; o^\ast \notin \mathcal{O}(c') \big].
% \]
\end{equation}
This metric isolates the primary attack objective independently of other caption changes.

\noindent \textbf{Global Preservation.}
Global Preservation measures how well non-target visual content is retained.
It is defined as the fraction of clean-caption objects that remain present after the attack:
\begin{equation}
% \[
\text{GlobalPreservation} =
\frac{|\mathcal{O}(c) \cap \mathcal{O}(c')|}
     {|\mathcal{O}(c)|}.
% \]
\end{equation}
A high preservation score indicates that the attack does not indiscriminately erase visual semantics.

\noindent \textbf{Grounded Hallucination Rate.}
Caption-level comparison alone cannot determine whether newly mentioned objects are visually supported.
To address this limitation, we introduce a \emph{grounded hallucination} metric that verifies adversarial caption objects against visual evidence using an external grounding model.

Let $\mathcal{H} = \mathcal{O}(c') \setminus \mathcal{O}(c)$ denote the set of newly introduced objects in the adversarial caption.
For each object $h \in \mathcal{H}$, we query a grounding model (GLIP \cite{li2022groundedlanguageimagepretraining})
% To determine whether newly mentioned objects are visually supported, we verify each object using a grounding model (GLIP).
% Specifically, an object mentioned in the adversarial caption is considered grounded only if GLIP produces a detection for that object in the image.
% Objects that cannot be grounded are counted as hallucinations.
%(e.g., GLIP, OWL-ViT, or GroundingDINO) 
to determine whether the object can be visually localized in the image.
Formally, let
\[
\text{detect}(h, I') =
\begin{cases}
1 & \text{if the grounding model localizes object } h \text{ with confidence } > \tau, \\
0 & \text{otherwise.}
\end{cases}
\]

Objects that cannot be grounded are considered hallucinated.
The grounded hallucination rate is defined as:
\begin{equation}
% \[
\text{GroundedHallucinationRate} =
\frac{|\{ h \in \mathcal{H} \;|\; \text{detect}(h, I') = 0 \}|}
     {|\mathcal{O}(c')|}.
% \]
\end{equation}

This formulation ensures that hallucination is measured relative to actual visual evidence rather than only caption differences.

\noindent \textbf{Head-Noun Grounded Hallucination.}
To account for paraphrasing and compound nouns (e.g., \emph{``salt shaker''} vs.\ \emph{``shaker''}),
we additionally report a head-noun grounded hallucination rate.
Each object phrase is reduced to its syntactic head noun prior to grounding verification, yielding a more semantically robust estimate.

\noindent \textbf{Semantic Drift.}
While hallucination focuses on unsupported object introduction, Semantic Drift measures global caption deviation.
Using a text embedding function $\phi(\cdot)$ (e.g., a sentence encoder), we compute: 
$\text{SemanticDrift} = 1 - \cos\big( \phi(c), \phi(c') \big).$
% \[
% \text{SemanticDrift} =
% 1 - \cos\big( \phi(c), \phi(c') \big).
% \]
This metric ensures that reduced hallucination is not achieved by collapsing captions into generic or uninformative descriptions.
Together, these metrics provide a fine-grained evaluation of concealment attacks.
% In contrast to prior work that relies primarily on caption similarity or attention statistics, our framework explicitly disentangles object removal, background preservation, and grounded hallucination.

% Importantly, the grounded hallucination metric verifies caption claims against visual evidence using an independent detection model, reducing the risk of self-referential evaluation.
% This distinction is critical when evaluating concealment attacks, where the goal is to remove visual evidence of a specific object without inducing unsupported object hallucinations.
Unlike CHAIR \cite{rohrbach2018object} and related hallucination metrics \cite{liu2024survey}, which assume static object presence and operate at coarse category granularity, our approach explicitly evaluates grounded object support and is designed for adversarial object concealment scenarios.
\section{Experimental Setup}
\label{sec:setup}

\subsection{Models}
To evaluate the effectiveness and generality of our attack, we consider three widely used vision--language models (VLMs) with distinct architectural designs: LLaVA-1.5 which combines a CLIP-based visual encoder with the Vicuna-7B large language model, BLIP-2 paired with the Flan-T5-XL language model via a Q-Former alignment module, and InstructBLIP which also employs a Vicuna-7B language model with instruction-tuned visual--language alignment.
% \begin{itemize}
%     \item \textbf{LLaVA-1.5}, which combines a CLIP-based visual encoder with the Vicuna-7B large language model,
%     \item \textbf{BLIP-2}, paired with the Flan-T5-XL language model via a Q-Former alignment module, and
%     \item \textbf{InstructBLIP}, which also employs a Vicuna-7B language model with instruction-tuned visual--language alignment.
% \end{itemize}
These models have been extensively adopted in prior work on adversarial attacks and robustness evaluation for VLMs~\cite{meftah2025vip, luo2024image, hu2024firm}.
They represent diverse design choices in terms of visual encoders, cross-modal fusion mechanisms, and decoding strategies.
For brevity, we refer to these models as LLaVA, BLIP2-T5, and InstructBLIP throughout the remainder of the paper.

\subsection{Datasets}
We evaluate our attack on two standard large-scale vision datasets with object-level annotations.
\noindent\textit{ImageNet:}
We randomly sample 1{,}000 images from the ImageNet validation set~\cite{russakovsky2015imagenet}.
For each image, we treat a ground-truth bounding box as the region of interest (ROI) and define the attack objective as preventing the VLM from detecting or describing the object within this region.
\noindent\textit{COCO:}
In addition, we evaluate our method on images from the COCO dataset \cite{caesar2018coco}, which contains complex multi-object scenes with diverse object categories and spatial layouts.
COCO allows us to assess the robustness of our attack under more challenging conditions, including overlapping objects and crowded scenes.
As with ImageNet, ground-truth bounding boxes are used to define ROIs, and the attack targets a single object category per image while preserving the remaining scene content.
% Across both datasets, images are resized to match the input resolution required by each model, and bounding boxes are rescaled accordingly.

% \subsection{Attack Protocol}
% Given a clean image $I$ and a target object $o^\ast$ with bounding box $\mathcal{B}$, we generate an adversarial image $I'$ by optimizing pixel values \emph{only within the ROI}.
% The attack operates under an $\ell_\infty$ perturbation budget in normalized image space and does not modify pixels outside the target region.
% For all experiments, we initialize the adversarial image from the clean input. Optimize for a fixed number of gradient steps using the Adam optimizer. Clamp perturbations to remain within the specified budget.
% % \begin{itemize}
% %     \item Initialize the adversarial image from the clean input.
% %     \item Optimize for a fixed number of gradient steps using the Adam optimizer.
% %     \item Clamp perturbations to remain within the specified budget.
% % \end{itemize}

% Unless otherwise stated, we use the same attack hyperparameters across all models and datasets to ensure consistency.

% \subsection{Implementation Details}
% We use $\epsilon = 0.2$, $\lambda_{\mathrm{stat}} = 1$, $\lambda_{\mathrm{dict}} = 2 $, $\lambda_{\mathrm{pres}} = 3$, $\lambda_{\mathrm{tv}} = 1e-3$, $\tau = 0.07$

% As for the target layers: for InstructBLIP $\{8,9,10,11\}$, for LLaVa $\{12, 14, 16, 18, 20, 22, 24\}$, for BLIP $\{8,9,10,11 \}$
\subsection{Implementation Details}
Unless otherwise stated, all experiments use a perturbation budget of $\epsilon = 0.2$.
The loss weights are set to
$\lambda_{\mathrm{stat}} = 1$,
$\lambda_{\mathrm{dict}} = 1$,
$\lambda_{\mathrm{pres}} = 1$,
and $\lambda_{\mathrm{tv}} = 10^{-3}$,
with a temperature parameter $\tau = 0.07$.
BCR is applied to later vision transformer layers.
Specifically, we optimize layers $\{22,23,24,25\}$ for Instruct-BLIP,
$\{21,22,23,24\}$ for LLaVA,
and $\{22,23,24,25\}$ for BLIP2-T5. For BLIP-2, we apply BCR losses to the frozen ViT-g/14 vision encoder, prior to the Q-Former bottleneck. We do not operate on Q-Former tokens, as they discard spatial correspondence. The choice of targeted layers is discussed in Appendix B.

% \subsection{Baselines}
\noindent \textbf{Baselines}
We compare our BCR attack against attention-suppression-based concealment methods, including VIP.
All baseline methods are implemented using their official or publicly available code and evaluated under the same threat model and perturbation budget.

% \subsection{Caption Generation and Prompts}
\noindent \textbf{Caption Generation and Prompts.}
For caption-based evaluation, we adopt the same prompts used in VIP, including: \texttt{``Describe this picture.''}
% \begin{center}
% \texttt{``Describe this picture.''}
% \end{center}
For binary verification tasks, we use object-specific yes/no prompts (e.g., \texttt{``Is there a \{object\} in the image?''}).
All captions are generated using greedy decoding with a fixed maximum token budget.

% \subsection{Evaluation Protocol}
\noindent \textbf{Evaluation Protocol.} Each attack is evaluated using the metrics introduced in Section~\ref{sec:metrics}, including Concealment Success, Global Preservation, Hallucination Rate, and Semantic Drift.
Results are reported as averages over all test images and target objects.
To isolate the effect of object concealment, all comparisons between clean and adversarial images are performed using identical prompts and decoding settings. Further details and algorithm are presented in Appendix A.

\section{Main Results}
We evaluate our Background-Consistent Re-encoding (BCR) attack against existing ROI-based suppression methods, with a particular focus on object concealment effectiveness, grounded hallucination behavior, and semantic consistency.
All results are averaged over the evaluation set described in Section~\ref{sec:setup}. Table~\ref{tab:multi_model_results} summarizes the main quantitative results across three vision--language models.
While both VIP and BCR achieve high object concealment rates, their failure modes differ substantially. VIP relies on suppressing attention and value flow from the ROI tokens, which frequently introduces a semantic gap in the visual representation.
This disruption weakens the visual grounding available to the language model. As a result, the model often compensates by generating plausible but visually unsupported objects, leading to a high \emph{grounded hallucination} rate and increased semantic drift.
In contrast, BCR achieves comparable concealment performance while substantially reducing grounded hallucinations.
Instead of suppressing ROI tokens, BCR re-encodes them to be statistically and semantically consistent with surrounding background tokens. This preserves representational continuity in the visual encoder and prevents the emergence of anomalous feature patterns that trigger compensatory language generation. The improvements are consistent across models and datasets.
For example, on ImageNet with Instruct-BLIP, BCR reduces grounded hallucination from $0.43$ (VIP) to $0.19$ while increasing global preservation from $0.57$ to $0.81$.
Similar trends are observed across BLIP2-T5 and LLaVA, as well as on the more complex COCO scenes.

These results suggest that hallucination under concealment attacks is closely tied to representational discontinuity.
By maintaining feature-level continuity rather than removing information outright, BCR enables effective object hiding while preserving coherent scene understanding.

\begin{table}[htp]
\caption{
Multi-model evaluation on ImageNet and COCO subsets.
\textbf{C}: Concealment success,
\textbf{GP}: Global Preservation,
\textbf{GH}: Grounded Hallucination rate (lower is better),
\textbf{SD}: Semantic Drift.
%Grounded hallucination is verified using an external grounding model to determine whether newly mentioned objects in adversarial captions are visually supported.
% BCR consistently achieves strong concealment while substantially reducing grounded hallucination and semantic drift across vision--language models and datasets.
}
\centering
\tiny
\begin{tabular}{llcccccccccccc}
\toprule
\textbf{Dataset}
& \textbf{Method}
& \multicolumn{4}{c}{\textbf{Instruct-BLIP}} 
& \multicolumn{4}{c}{\textbf{BLIP2-T5}} 
& \multicolumn{4}{c}{\textbf{LLaVA}} \\
\cmidrule(lr){3-6} \cmidrule(lr){7-10} \cmidrule(lr){11-14}

&
& \textbf{C} $\uparrow$ & \textbf{GP} $\uparrow$ & \textbf{GH} $\downarrow$ & \textbf{SD} $\downarrow$
& \textbf{C} $\uparrow$ & \textbf{GP} $\uparrow$ & \textbf{GH} $\downarrow$ & \textbf{SD} $\downarrow$
& \textbf{C} $\uparrow$ & \textbf{GP} $\uparrow$ & \textbf{GH} $\downarrow$ & \textbf{SD} $\downarrow$ \\
\midrule

\multirow{5}{*}{\textbf{ImageNet}}
& No Attack
& 0.00 & 1.00 & 0.00 & 0.00
& 0.00 & 1.00 & 0.00 & 0.00
& 0.00 & 1.00 & 0.00 & 0.00 \\

& Masking
& 1.00 & 0.41 & 0.59 & 0.54
& 1.00 & 0.52 & 0.48 & 0.40
& 1.00 & 0.55 & 0.45 & 0.41 \\

& PRM
& 0.66 & 0.38 & 0.62 & 0.44
& 0.70 & 0.31 & 0.69 & 0.41
& 0.73 & 0.34 & 0.66 & 0.39 \\

& VIP 
& 0.93 & 0.57 & 0.43 & 0.35 
& 0.96 & 0.59 & 0.41 & 0.32
& 0.98 & 0.62 & 0.48 & 0.29 \\

& \textbf{BCR (Ours)}
& \textbf{0.92} & \textbf{0.81} & \textbf{0.19} & \textbf{0.13} 
& \textbf{0.95} & \textbf{0.83} & \textbf{0.17} & \textbf{0.10}
& \textbf{0.98} & \textbf{0.86} & \textbf{0.14} & \textbf{0.08} \\

\midrule

\multirow{5}{*}{\textbf{COCO}}
& No Attack
& 0.00 & 1.00 & 0.00 & 0.00
& 0.00 & 1.00 & 0.00 & 0.00
& 0.00 & 1.00 & 0.00 & 0.00 \\

& Masking
& 1.00 & 0.37 & 0.63 & 0.46
& 1.00 & 0.39 & 0.61 & 0.44
& 1.00 & 0.41 & 0.59 & 0.41 \\

& PRM
& 0.63 & 0.54 & 0.46 & 0.37
& 0.77 & 0.58 & 0.42 & 0.34
& 0.80 & 0.60 & 0.40 & 0.33 \\

& VIP 
& 0.80 & 0.42 & 0.58 & 0.48 
& 0.83 & 0.45 & 0.55 & 0.45
& 0.88 & 0.48 & 0.52 & 0.43 \\

& \textbf{BCR (Ours)}
& \textbf{0.89} & \textbf{0.77} & \textbf{0.23} & \textbf{0.16} 
& \textbf{0.92} & \textbf{0.79} & \textbf{0.21} & \textbf{0.13}
& \textbf{0.95} & \textbf{0.82} & \textbf{0.18} & \textbf{0.11} \\

\bottomrule
\end{tabular}
\label{tab:multi_model_results}
\end{table}
\subsection{Qualitative Comparison with Attention Suppression}
%%===================================%%

\begin{figure*}[htp]
    \centering
    \includegraphics[width=\linewidth]{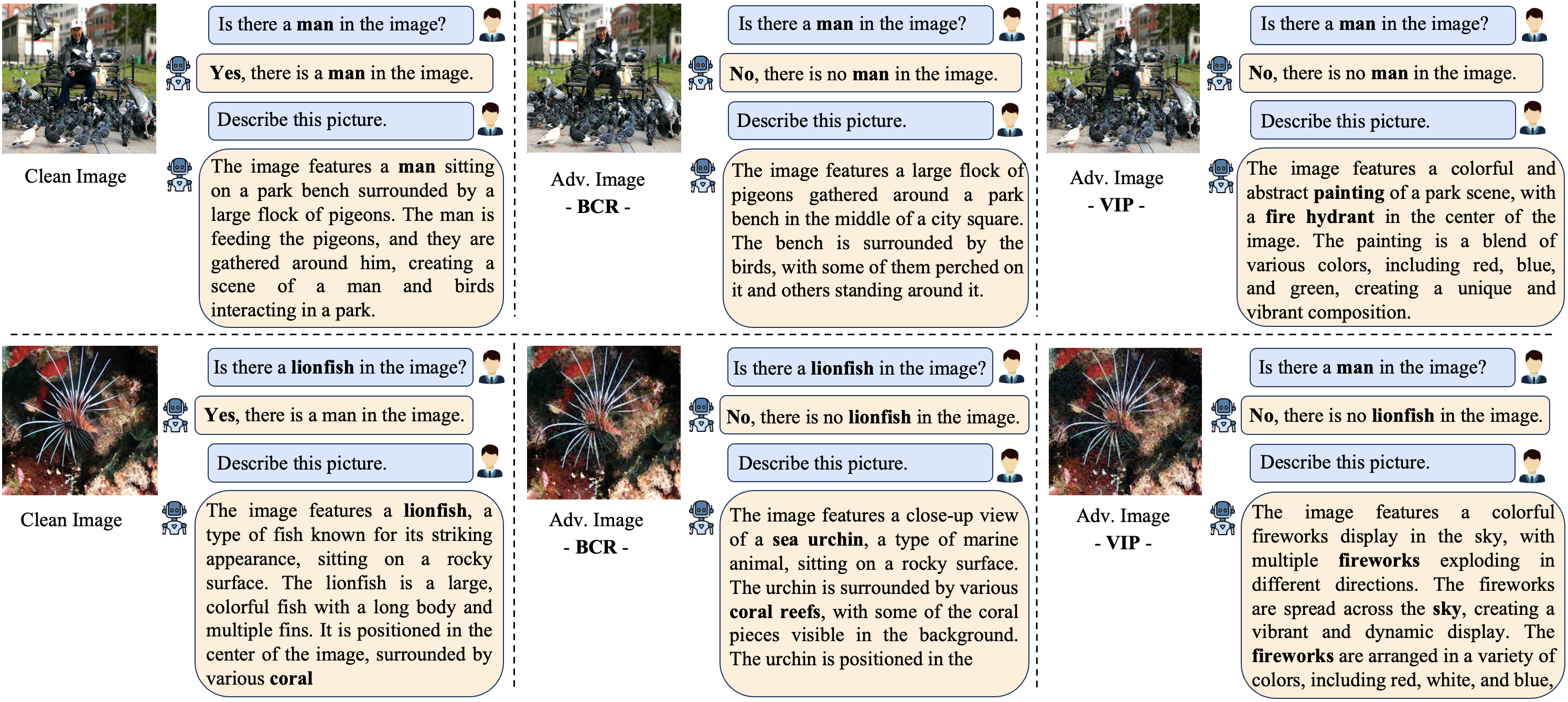}
    \caption{
    Qualitative comparison of object concealment attacks on InstructBLIP.
    % For each example, the clean caption is compared with captions generated after applying VIP and BCR.
    %While both methods successfully conceal the target object, VIP frequently produces hallucinated or semantically unrelated descriptions. In contrast, BCR preserves scene context and generates captions that remain visually grounded after concealment.
    }
    \label{fig:qualitative_res}
\end{figure*}

We present qualitative comparisons between our Background-Consistent Re-encoding (BCR) attack and attention-suppression-based methods such as VIP.
Figure~\ref{fig:qualitative_res} shows representative examples where the target object lies within the annotated region of interest (ROI).
For each case, we report the clean caption together with captions generated after applying VIP and BCR.

In the first example, both methods successfully conceal the target person.
However, VIP substantially alters the global scene description, introducing unrelated objects such as a \emph{fire hydrant} and describing the scene as an abstract painting.
This behavior reflects a common failure mode of attention suppression: by removing the contribution of ROI tokens, the visual representation becomes incomplete, prompting the language model to hallucinate visually unsupported content.
In contrast, BCR removes the person while preserving the surrounding visual context.
The generated caption continues to reference grounded elements of the scene, such as the pigeons and park bench, without introducing spurious objects.
The second example highlights a different failure mode.
VIP again removes the target object but produces a caption describing an unrelated fireworks scene, which bears little resemblance to the underlying image.
This reflects severe semantic drift caused by the disruption of the visual representation.
In contrast, BCR replaces the concealed object with a visually plausible marine entity (a sea urchin).
Although this represents a semantic substitution rather than a strict omission, the resulting description remains consistent with the visual environment (coral reef and marine background) and does not introduce unrelated objects. Across examples, attention suppression often creates a representational void that the language model compensates for by generating plausible but visually unsupported content.
BCR mitigates this effect by preserving representational continuity within the visual encoder.
Instead of removing ROI tokens, BCR re-encodes them to align with the statistical structure of surrounding background tokens.
This prevents anomalous feature patterns that would otherwise trigger compensatory hallucination during language generation.
Overall, BCR consistently conceals the target object while maintaining coherent scene descriptions, favoring contextually grounded substitutions over hallucinated or unrelated content.
Additional qualitative results are provided in Appendix~C.
% %%===============================
% \subsection{Comparison with Pixel-Space Obfuscation Baselines}
% %%===============================

% Beyond attention-suppression attacks, a natural class of object concealment baselines operates directly in pixel space, including \emph{blurring}, \emph{masking}, and \emph{inpainting} of the target region.
% These methods are widely used for visual privacy and anonymization and are often assumed to reduce hallucination by removing explicit object evidence.
% We qualitatively compare these baselines with BCR to highlight fundamental differences in their downstream effects on vision--language models.
% Figure \ref{fig:mask} illustrates representative failure cases.
% When the target region is blurred or masked, the resulting image contains visually unnatural or ambiguous artifacts.
% Despite the absence of explicit object cues, VLMs frequently hallucinate new entities to explain these artifacts.
% For example, in a scene originally depicting chickens near a fence, pixel-space obfuscation causes the model to hallucinate multiple dogs and a field, introducing entirely new object categories that were never present.
% This behavior mirrors the hallucination observed under attention suppression and indicates that visual ambiguity alone is sufficient to trigger semantic fabrication.

% \begin{figure*}[h!]
%     \centering
%     \includegraphics[width=\linewidth]{figures/mask_.png}
%     \caption{Qualitative results of pixel-space obfuscation}
%     \label{fig:mask}
% \end{figure*}
%%===============================
\subsection{Comparison with Pixel-Space Obfuscation Baselines}
\begin{figure*}[htp]
    \centering
    \includegraphics[width=\linewidth]{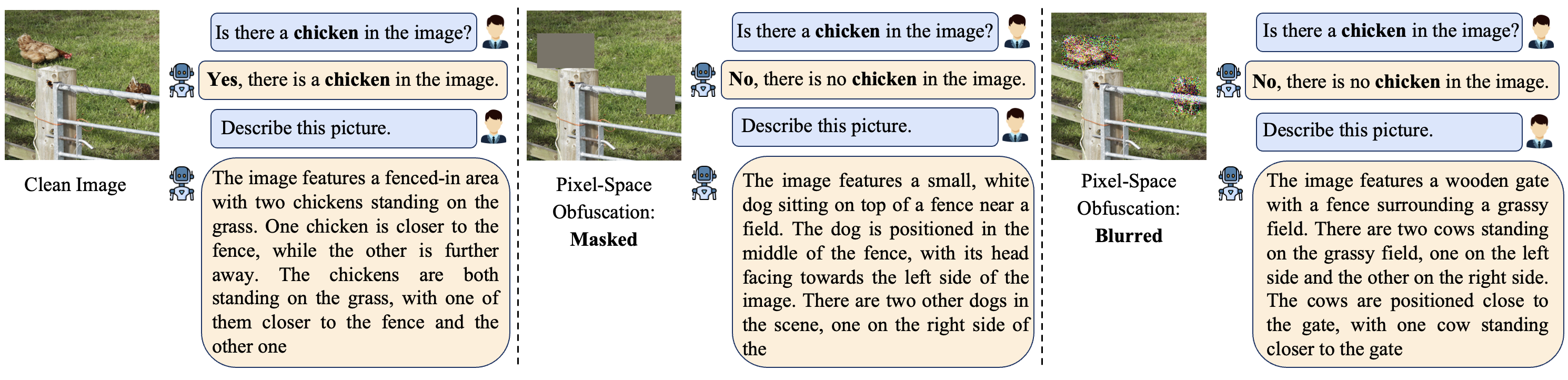}
    \caption{
    Failure cases of pixel-space obfuscation methods.
    When the target object is masked or blurred, the resulting visual artifacts introduce ambiguous signals that vision--language models attempt to explain.
    Despite the absence of the original object, the model generates hallucinated descriptions containing unrelated entities such as dogs or cows.
    % These results illustrate that removing visual evidence alone can increase semantic ambiguity and lead to hallucinated scene interpretations.
    }
    \label{fig:mask}
\end{figure*}

Beyond attention-suppression attacks, another class of concealment baselines operates directly in pixel space, including \emph{blurring}, \emph{masking}, and \emph{inpainting} of the target region.
These techniques are widely used in visual privacy and anonymization, as they remove explicit visual evidence of the target object.
However, their interaction with vision--language models can introduce unintended semantic artifacts.
Figure~\ref{fig:mask} shows representative examples.
When the target region is blurred or masked, the resulting image often contains ambiguous or unnatural patterns that the vision encoder struggles to interpret.
As a result, the language model attempts to explain these signals by generating plausible but visually unsupported descriptions.
For example, in a scene originally depicting chickens near a fence, pixel-space obfuscation leads the model to hallucinate unrelated objects such as dogs or cows.
This illustrates a key limitation of pixel-space concealment: removing visual evidence alone can introduce ambiguity that triggers compensatory reasoning in the language model, resulting in hallucinated scene interpretations.
% \paragraph{Semantic Substitution vs. Hallucination.}
% In contrast, BCR produces adversarial images that preserve low-level continuity and global scene structure.
% As a result, captions generated under BCR rarely introduce out-of-context objects.
% Instead, when concealment is imperfect, BCR tends to induce \emph{semantic substitution}, where the hidden object is replaced by a contextually plausible alternative.
% For instance, concealing a flute held by a person may result in the model describing a piano rather than inventing an unrelated object.
% While this substitution reflects a failure to fully eliminate object semantics, it preserves scene coherence and avoids the more disruptive hallucination failures observed in pixel-space baselines.

% These observations suggest that hallucination in VLMs is not primarily caused by object absence, but by \emph{representational discontinuity and visual ambiguity}.
% Pixel-space obfuscation removes evidence but introduces artifacts that violate the model’s learned visual statistics, prompting compensatory hallucination.
% By maintaining representational continuity, BCR shifts failure modes from hallucination toward semantic substitution, which we view as a strictly weaker and more interpretable error.
% This distinction motivates future work on combining background-consistent re-encoding with stronger semantic disentanglement to further suppress residual object identity.
%%==============================
\subsection{Perceptual Fidelity}
%%==============================
In addition to semantic preservation, effective object concealment should maintain high visual fidelity so that perturbations remain unobtrusive.
We therefore measure perceptual similarity between clean and adversarial images using Structural Similarity (SSIM) and Learned Perceptual Image Patch Similarity (LPIPS).
Table~\ref{tab:perceptual_fidelity} shows that BCR consistently achieves higher SSIM and lower LPIPS across models compared to attention-suppression attacks.
\begin{wraptable}{r}{7.5cm}
\caption{Perceptual fidelity between clean and adversarial images. Higher SSIM and lower LPIPS indicate better visual similarity.}
\centering
\tiny
\begin{tabular}{lcccccc}
\toprule
 \textbf{Model}
& \multicolumn{2}{c}{\textbf{Instruct-BLIP}} 
& \multicolumn{2}{c}{\textbf{BLIP2-T5}} 
& \multicolumn{2}{c}{\textbf{LLaVA}} \\
\cmidrule(r){2-3} \cmidrule(lr){4-5} \cmidrule(lr){6-7}

\textbf{Method} 
&  \textbf{SSIM} $\uparrow$ & \textbf{LPIPS} $\downarrow$
& \textbf{SSIM} $\uparrow$ & \textbf{LPIPS} $\downarrow$
& \textbf{SSIM} $\uparrow$ & \textbf{LPIPS} $\downarrow$ \\
\midrule

PRM 
& 0.93 & 0.06
& 0.93 & 0.06
& 0.92 & 0.07 \\

VIP 
& 0.71 & 0.34
& 0.804 & 0.175
& 0.65 & 0.32 \\

\textbf{BCR (Ours)}
& \textbf{0.96} & \textbf{0.02}
& \textbf{0.95} & \textbf{0.03}
& \textbf{0.94} & \textbf{0.04} \\

\bottomrule
\end{tabular}
\label{tab:perceptual_fidelity}
\end{wraptable}
Although BCR modifies pixels within the ROI, the perturbations remain visually coherent with surrounding content, avoiding the artifacts commonly introduced by suppression-based methods.
These results indicate that BCR preserves both semantic consistency and perceptual realism, producing adversarial images that remain visually close to the original inputs.

\section{Ablation Study}
\begin{wraptable}{r}{5cm}
\centering
\tiny
\caption{Ablation study of BCR components on Instruct-BLIP.
$\checkmark$ indicates the loss term is enabled.
We report Concealment Success (CS~$\uparrow$), Hallucination Rate (HR~$\downarrow$),
and Global Preservation (GP~$\uparrow$).}
\label{tab:ablation}
\begin{tabular}{c c c c | c c c}
\toprule
$\mathcal{L}_{\text{stat}}$ &
$\mathcal{L}_{\text{dict}}$ &
$\mathcal{L}_{\text{pres}}$ &
$\mathcal{L}_{\text{tv}}$ &
CS $\uparrow$ &
HR $\downarrow$ &
GP $\uparrow$ \\
\midrule
\checkmark & \checkmark & \checkmark & \checkmark & \textbf{0.91} & \textbf{0.08} & \textbf{0.83} \\
\checkmark & \checkmark & \checkmark &            & 0.90 & 0.09 & 0.81 \\
\checkmark & \checkmark &            & \checkmark & 0.86 & 0.17 & 0.69 \\
\checkmark &            & \checkmark & \checkmark & 0.79 & 0.21 & 0.74 \\
            & \checkmark & \checkmark & \checkmark & 0.72 & 0.28 & 0.66 \\
\midrule
\multicolumn{4}{c|}{VIP-style suppression baseline} & 0.88 & 0.42 & 0.51 \\
\bottomrule
\end{tabular}
\end{wraptable}
We conduct an ablation study to isolate the contribution of each component in BCR. All variants are evaluated on the same images and ROIs with identical optimization and evaluation settings. Table~\ref{tab:ablation} shows that the full objective achieves the best balance between concealment success and semantic stability.
Removing either the statistical alignment loss ($\mathcal{L}_{\text{stat}}$) or the dictionary projection loss ($\mathcal{L}_{\text{dict}}$) substantially increases hallucination, indicating that both distributional matching and semantic anchoring are necessary for stable representations.
% Unless stated otherwise, results are reported on Instruct-BLIP; similar trends are observed for LLaVA and BLIP2-T5.
Similarly, disabling background preservation ($\mathcal{L}_{\text{pres}}$) degrades global semantics, reducing GP scores.
Overall, the results confirm that BCR’s effectiveness stems from enforcing representational continuity at both statistical and semantic levels.

\section{Conclusion}

% We presented Background-Consistent Re-encoding (BCR), a continuity-preserving adversarial attack for vision--language models that conceals target objects without inducing hallucination.
% Unlike suppression-based attacks that remove visual evidence and create representational gaps, BCR re-encodes the target region to remain statistically and semantically consistent with its surrounding context, thereby maintaining global coherence.
% Our findings reveal a strong link between hallucination and representational discontinuity in multimodal models, suggesting that continuity-aware manipulation is a more principled alternative to suppression.
% We believe this perspective opens new directions for adversarial evaluation, privacy protection, and robust multimodal perception.
We introduced Background-Consistent Re-encoding (BCR), a continuity-preserving attack that conceals target objects in vision--language models without inducing hallucination.
By re-encoding ROI features to align with surrounding context, BCR maintains representational continuity and scene coherence.
Our results suggest that hallucination arises largely from representational discontinuities rather than object absence itself.
This insight highlights continuity-aware manipulation as a promising direction for adversarial analysis and robust multimodal perception.

\clearpage  % TODO FINAL: This \clearpage needs to be removed from both review and camera-ready versions.

% \section*{Acknowledgements}
% Please insert your acknowledgments here.

\bibliographystyle{splncs04}
\bibliography{main}

@inproceedings{radford2021clip,
  title={Learning transferable visual models from natural language supervision},
  author={Radford, Alec and Kim, Jong Wook and Hallacy, Chris and Ramesh, Aditya and Goh, Gabriel and Agarwal, Sandhini and Sastry, Girish and Askell, Amanda and Mishkin, Pamela and Clark, Jack and others},
  booktitle={International conference on machine learning},
  pages={8748--8763},
  year={2021},
  organization={PmLR}
}

@article{alayrac2022flamingo,
  title={Flamingo: a visual language model for few-shot learning},
  author={Alayrac, Jean-Baptiste and Donahue, Jeff and Luc, Pauline and Miech, Antoine and Barr, Iain and Hasson, Yana and Lenc, Karel and Mensch, Arthur and Millican, Katherine and Reynolds, Malcolm and others},
  journal={Advances in neural information processing systems},
  volume={35},
  pages={23716--23736},
  year={2022}
}

@inproceedings{li2023blip2,
  title={Blip-2: Bootstrapping language-image pre-training with frozen image encoders and large language models},
  author={Li, Junnan and Li, Dongxu and Savarese, Silvio and Hoi, Steven},
  booktitle={International conference on machine learning},
  pages={19730--19742},
  year={2023},
  organization={PMLR}
}

@article{dai2023instructblip,
  title={Instructblip: Towards general-purpose vision-language models with instruction tuning},
  author={Dai, Wenliang and Li, Junnan and Li, Dongxu and Tiong, Anthony and Zhao, Junqi and Wang, Weisheng and Li, Boyang and Fung, Pascale N and Hoi, Steven},
  journal={Advances in neural information processing systems},
  volume={36},
  pages={49250--49267},
  year={2023}
}

@article{zhang2024visual,
  title={Visual adversarial attack on vision-language models for autonomous driving},
  author={Zhang, Tianyuan and Wang, Lu and Zhang, Xinwei and Zhang, Yitong and Jia, Boyi and Liang, Siyuan and Hu, Shengshan and Fu, Qiang and Liu, Aishan and Liu, Xianglong},
  journal={arXiv preprint arXiv:2411.18275},
  year={2024}
}

@misc{li2022groundedlanguageimagepretraining,
      title={Grounded Language-Image Pre-training}, 
      author={Liunian Harold Li and Pengchuan Zhang and Haotian Zhang and Jianwei Yang and Chunyuan Li and Yiwu Zhong and Lijuan Wang and Lu Yuan and Lei Zhang and Jenq-Neng Hwang and Kai-Wei Chang and Jianfeng Gao},
      year={2022},
      eprint={2112.03857},
      archivePrefix={arXiv},
      primaryClass={cs.CV},
      url={https://arxiv.org/abs/2112.03857}, 
}

@article{liu2024survey,
  title={A survey on hallucination in large vision-language models},
  author={Liu, Hanchao and Xue, Wenyuan and Chen, Yifei and Chen, Dapeng and Zhao, Xiutian and Wang, Ke and Hou, Liping and Li, Rongjun and Peng, Wei},
  journal={arXiv preprint arXiv:2402.00253},
  year={2024}
}

@article{datta2025evaluating,
  title={Evaluating Hallucination in Large Vision-Language Models based on Context-Aware Object Similarities},
  author={Datta, Shounak and Sundararaman, Dhanasekar},
  journal={arXiv preprint arXiv:2501.15046},
  year={2025}
}

@article{goodfellow2014explaining,
  title={Explaining and harnessing adversarial examples},
  author={Goodfellow, Ian J and Shlens, Jonathon and Szegedy, Christian},
  journal={arXiv preprint arXiv:1412.6572},
  year={2014}
}

@article{madry2017towards,
  title={Towards deep learning models resistant to adversarial attacks},
  author={Madry, Aleksander and Makelov, Aleksandar and Schmidt, Ludwig and Tsipras, Dimitris and Vladu, Adrian},
  journal={arXiv preprint arXiv:1706.06083},
  year={2017}
}

@article{meftah2025vip,
  title={VIP: Visual Information Protection through Adversarial Attacks on Vision-Language Models},
  author={Meftah, Hanene FZ and Hamidouche, Wassim and Fezza, Sid Ahmed and D{\'e}forges, Olivier},
  journal={arXiv preprint arXiv:2507.08982},
  year={2025}
}

@article{luo2024image,
  title={An image is worth 1000 lies: Adversarial transferability across prompts on vision-language models},
  author={Luo, Haochen and Gu, Jindong and Liu, Fengyuan and Torr, Philip},
  journal={arXiv preprint arXiv:2403.09766},
  year={2024}
}

@article{hu2024firm,
  title={As Firm As Their Foundations: Can open-sourced foundation models be used to create adversarial examples for downstream tasks?},
  author={Hu, Anjun and Gu, Jindong and Pinto, Francesco and Kamnitsas, Konstantinos and Torr, Philip},
  journal={arXiv preprint arXiv:2403.12693},
  year={2024}
}

@article{rohrbach2018object,
  title={Object hallucination in image captioning},
  author={Rohrbach, Anna and Hendricks, Lisa Anne and Burns, Kaylee and Darrell, Trevor and Saenko, Kate},
  journal={arXiv preprint arXiv:1809.02156},
  year={2018}
}

@article{russakovsky2015imagenet,
  title={Imagenet large scale visual recognition challenge},
  author={Russakovsky, Olga and Deng, Jia and Su, Hao and Krause, Jonathan and Satheesh, Sanjeev and Ma, Sean and Huang, Zhiheng and Karpathy, Andrej and Khosla, Aditya and Bernstein, Michael and others},
  journal={International journal of computer vision},
  volume={115},
  number={3},
  pages={211--252},
  year={2015},
  publisher={Springer}
}

@inproceedings{caesar2018coco,
  title={Coco-stuff: Thing and stuff classes in context},
  author={Caesar, Holger and Uijlings, Jasper and Ferrari, Vittorio},
  booktitle={Proceedings of the IEEE conference on computer vision and pattern recognition},
  pages={1209--1218},
  year={2018}
}

% Appendix A:
% BCR algorithm
% implementation details

% Appendix B:
% The choice of targeted layers

% Appendix C:
% further results

\section*{A. Additional Implentation Details}
\subsection*{A.1. Algorithm Overview}

Algorithm~\ref{alg:bcr} summarizes the proposed Background-Consistent Re-encoding (BCR) attack.
Given an input image and a region of interest (ROI), the algorithm optimizes a pixel-level adversarial image under a bounded perturbation budget.
At each iteration, visual features are extracted from selected layers of the vision encoder.
ROI token representations are then encouraged to (i) match the first- and second-order statistics of background tokens, (ii) lie within the background feature manifold via soft dictionary projection, and (iii) preserve non-ROI features to maintain global scene semantics.
A total variation regularizer enforces spatial smoothness of the perturbation.
The optimization proceeds via gradient descent in pixel space, producing an adversarial image that conceals the target object while preserving semantic continuity and reducing hallucination.

\begin{algorithm} %[t]
\caption{Background-Consistent Re-encoding (BCR)}
\label{alg:bcr}
\small
\begin{algorithmic}[1]

\REQUIRE Image $\mathbf{x}$, ROI boxes $\mathcal{B}$, vision encoder $f_\theta$, layers $\mathcal{L}$,
step size $\eta$, steps $T$, perturbation budget $\epsilon$
\ENSURE Adversarial image $\mathbf{x}^{adv}$

\STATE Compute ROI pixel mask $\mathbf{M}$ from $\mathcal{B}$
\STATE Compute ROI token indices $\mathcal{I}_r$ and background indices $\mathcal{I}_b$
\STATE Initialize $\mathbf{x}^{adv} \leftarrow \mathbf{x}$

\FOR{each layer $l \in \mathcal{L}$}
    \STATE Store clean features $\mathbf{Z}^{(l)}(\mathbf{x})$
\ENDFOR

\FOR{$t = 1$ to $T$}
    \FOR{each layer $l \in \mathcal{L}$}
        \STATE Extract ROI features $\mathbf{Z}^{(l)}_r$ and background features $\mathbf{Z}^{(l)}_b$
        \STATE Compute statistical loss $\mathcal{L}^{(l)}_{\mathrm{stat}}$
        \STATE Compute dictionary projection loss $\mathcal{L}^{(l)}_{\mathrm{dict}}$
        \STATE Compute background preservation loss $\mathcal{L}^{(l)}_{\mathrm{pres}}$
    \ENDFOR

    \STATE Compute total variation loss $\mathcal{L}_{\mathrm{tv}}$ on ROI
    \STATE $\mathcal{L} \leftarrow \sum_{l \in \mathcal{L}}
        \big(
        \lambda_{\mathrm{stat}}\mathcal{L}^{(l)}_{\mathrm{stat}}
        + \lambda_{\mathrm{dict}}\mathcal{L}^{(l)}_{\mathrm{dict}}
        + \lambda_{\mathrm{pres}}\mathcal{L}^{(l)}_{\mathrm{pres}}
        \big)
        + \lambda_{\mathrm{tv}}\mathcal{L}_{\mathrm{tv}}$

    \STATE Update $\mathbf{x}^{adv}$ using gradient descent on $\mathcal{L}$
    \STATE Project perturbation to $\ell_\infty$ budget $\epsilon$
\ENDFOR

% \RETURN $\mathbf{x}^{adv}$
\STATE \textbf{return} $\mathbf{x}^{adv}$

\end{algorithmic}
\end{algorithm}

\subsection*{A.2. Grounded Hallucination Verification with GLIP}
\label{sec:glip_grounding}

To ensure that hallucination measurements reflect the presence of visually unsupported objects rather than linguistic variation, we augment our caption-based evaluation with a grounding-based verification step.

% \paragraph{Motivation.}
Caption comparison alone can overestimate hallucination due to paraphrasing or synonym usage. For example, an object may appear in a caption under a different lexical form (e.g., ``automobile'' vs.\ ``car''). Conversely, models may introduce object terms that are not visually supported in the image. To disambiguate these cases, we verify whether newly mentioned objects can be grounded in the image.

% \paragraph{Object Extraction.}
Given a clean image $I$ and its adversarial counterpart $I'$, we generate captions $c = f(I)$ and $c' = f(I')$ using the same prompt and decoding settings. Object candidates are extracted from both captions using a dependency-based noun phrase parser. After lemmatization and normalization, we obtain the object sets
\[
O(c), \quad O(c').
\]

The set of newly introduced objects is defined as
\[
H_{\text{cand}} = O(c') \setminus O(c).
\]

% \paragraph{GLIP-Based Grounding.}
For each candidate object $o \in H_{\text{cand}}$, we query the Grounded Language–Image Pretraining (GLIP) detector to verify whether the object can be localized in the adversarial image $I'$. GLIP predicts bounding boxes conditioned on the textual query corresponding to the object name. If no detection with confidence above a predefined threshold $\tau_g$ is produced, the object is considered visually unsupported.

Formally, the hallucination set is defined as
\[
H = \{\, o \in H_{\text{cand}} \mid \text{GLIP}(I', o) = \varnothing \,\}.
\]

% \paragraph{Grounded Hallucination Rate.}
Using the verified hallucination set $H$, we compute the grounded hallucination rate as
\[
\text{HallucinationRate}_{\text{grounded}} =
\frac{|H|}{|O(c')|}.
\]

This metric measures the fraction of objects mentioned in the adversarial caption that cannot be visually grounded in the image.

% \paragraph{Implementation Details.}
We use a publicly available GLIP model pretrained on large-scale grounding datasets. The grounding confidence threshold is set to $\tau_g = 0.3$. For multi-word phrases, the head noun is used as the textual query to improve grounding robustness.
% \paragraph{Discussion.}
This grounding-based verification reduces ambiguity in hallucination evaluation by ensuring that newly introduced object tokens correspond to detectable visual evidence. By combining caption-based object extraction with grounding verification, the evaluation more accurately captures hallucinations caused by adversarial manipulation rather than lexical variation.

\section*{B. Sensitivity Analysis of Targeted Transformer Layers}
\label{sec:layer_sensitivity}

Our Background-Consistent Re-encoding (BCR) objective enforces statistical and semantic alignment between region-of-interest (ROI) tokens and background tokens across a subset of vision transformer layers. The choice of targeted layers influences the strength and stability of the concealment attack, as different layers encode visual information at varying levels of abstraction.

Early layers of vision transformers primarily capture low-level features such as edges, textures, and local patterns, while deeper layers encode higher-level semantic representations that are more directly aligned with language generation. Consequently, applying the BCR objective at different depths may affect the degree to which object semantics are suppressed or redistributed into background representations.

\begin{figure*}[htp]
    \centering
    \includegraphics[width=\linewidth]{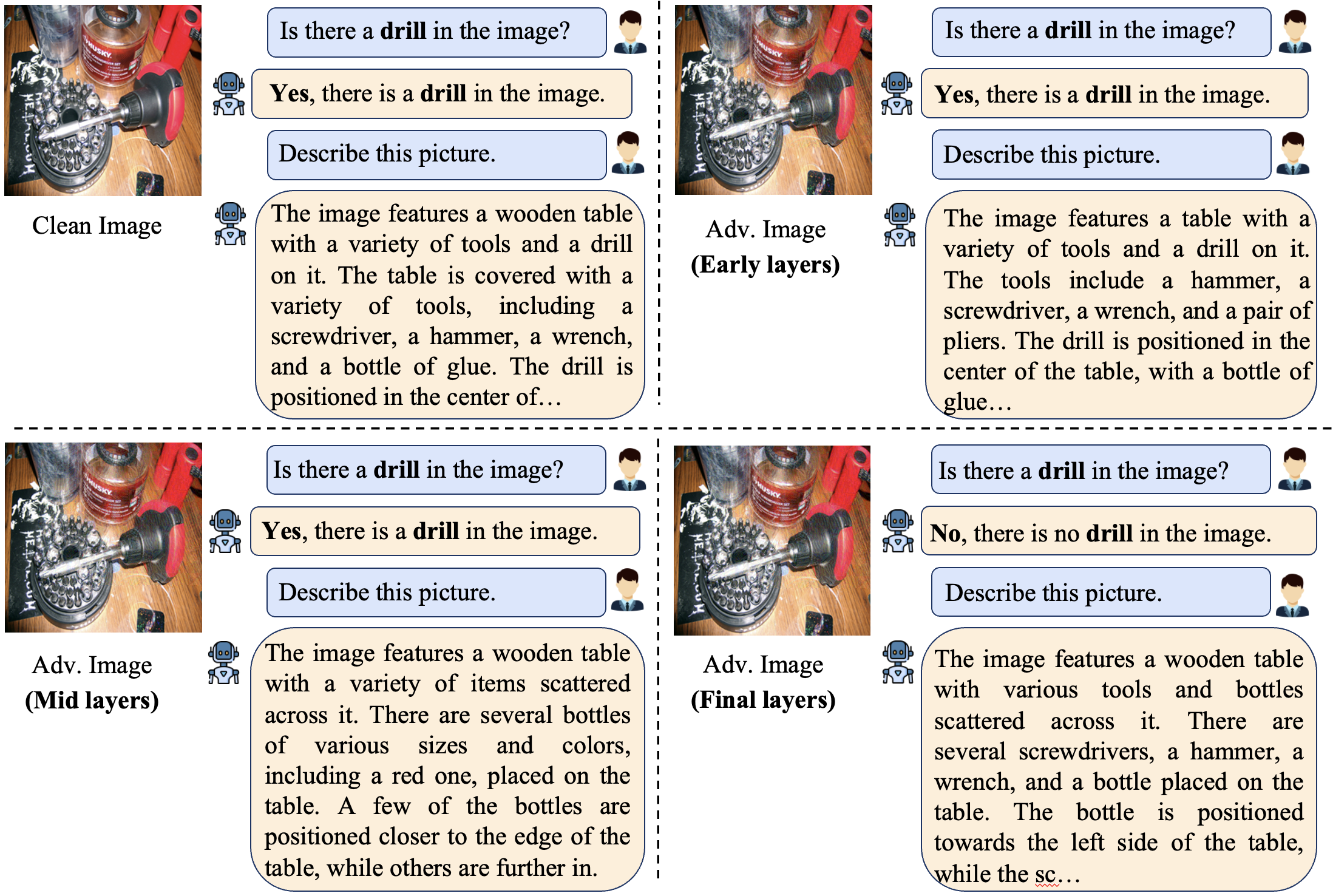}
    \caption{Sensitivity analysis of targeted transformer layers. Early-layer optimization leaves object semantics largely intact, allowing the model to still recognize the target object. In contrast, targeting deeper layers removes the object semantics while preserving the surrounding scene context, resulting in successful concealment.}
    \label{fig:exp1}
\end{figure*}

To study this effect, we evaluate BCR by applying the alignment losses to different groups of transformer layers:

\begin{itemize}
\item \textbf{Early layers}: losses applied to the first four transformer blocks.
\item \textbf{Middle layers}: losses applied to intermediate transformer blocks.
\item \textbf{Late layers}: losses applied to the final four transformer blocks.
\end{itemize}

All other hyperparameters remain identical to the main experiments, including the perturbation budget and optimization schedule.

We additionally measured concealment success and hallucination rate across layer configurations, confirming that targeting deeper layers yields the most stable concealment behavior. Early-layer optimization leaves object semantics intact, allowing the model to still recognize the target object. In contrast, targeting deeper layers removes the object semantics while preserving the surrounding scene context.

\begin{table}[h]
\centering
\small
\caption{Quantitative sensitivity analysis of targeted transformer layers on Instruct-BLIP. Later-layer optimization yields stronger concealment and lower hallucination.}
\begin{tabular}{lcc}
\toprule
\textbf{Targeted Layers} & \textbf{Concealment Success $\uparrow$} & \textbf{Hallucination Rate $\downarrow$} \\
\midrule
Early Layers & 0.22 & 0.07 \\
Middle Layers & 0.41 & 0.11 \\
Late Layers & \textbf{0.91} & \textbf{0.18} \\
\bottomrule
\end{tabular}
\label{tab:layer_sensitivity}
\end{table}

These results indicate that hallucination and object recognition in vision--language models are primarily driven by higher-level semantic representations. By aligning ROI features with background features in deeper layers, BCR effectively removes object-specific semantics while preserving global scene coherence. Based on this analysis, all experiments in the main paper apply the BCR objective to later transformer layers of the vision encoder.

\section*{C. Additional Qualitative Results}

\begin{figure*}[htp]
    \centering
    \includegraphics[width=\linewidth]{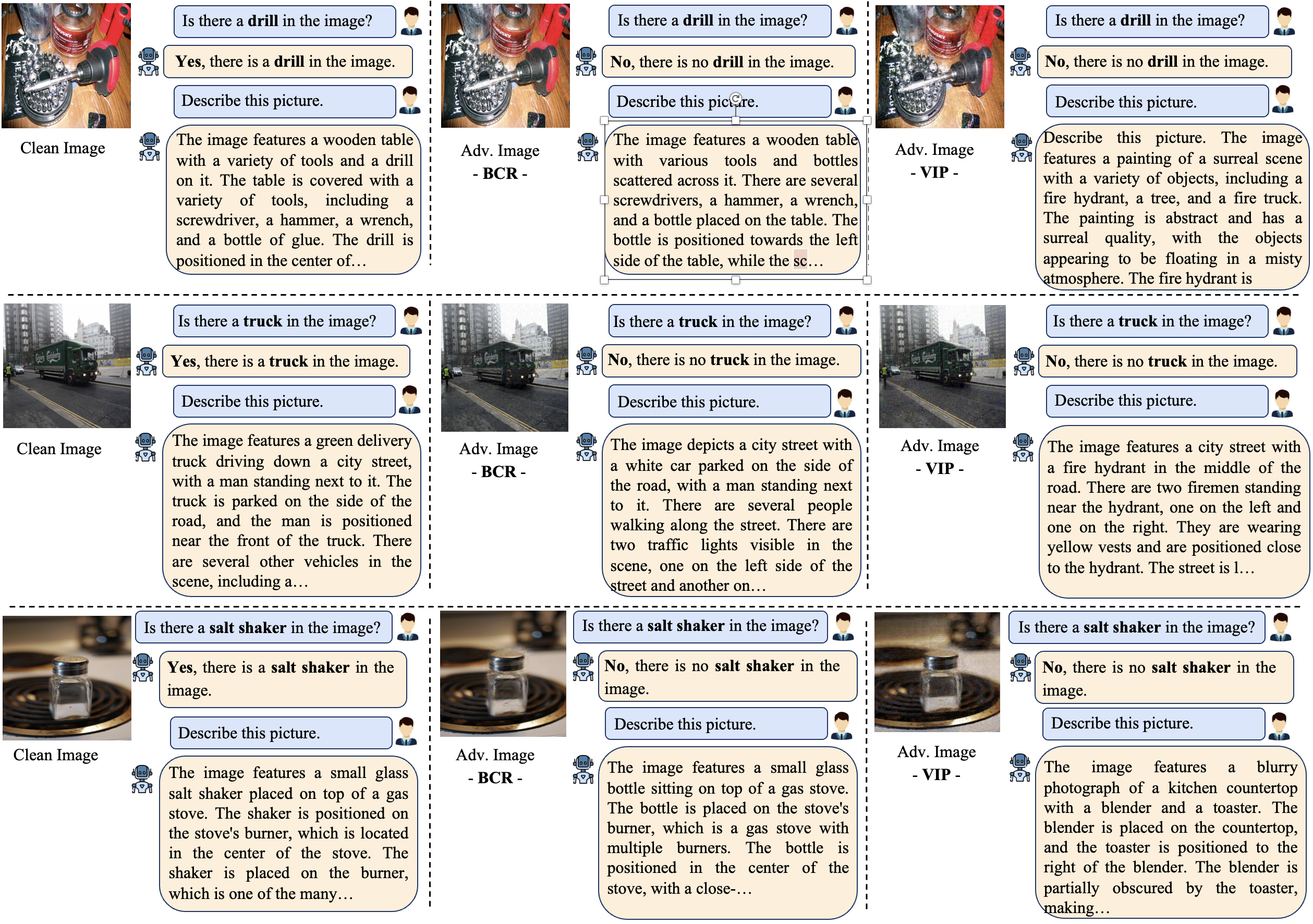}
    \caption{
Additional qualitative comparisons between suppression-based attacks and BCR.
While suppression-based methods often introduce hallucinated or semantically unrelated objects,
BCR consistently removes the target object while preserving the overall scene structure and contextual elements.
}
    \label{fig:exp2}
\end{figure*}

We present additional qualitative examples illustrating the behavior of BCR across different scenes and target objects (see Figure \ref{fig:exp2}).
For each example, we compare captions generated from the clean image and the adversarial images produced by different concealment strategies.

Across diverse scenes, suppression-based approaches frequently produce unstable descriptions.
Because these methods remove or weaken ROI representations, the resulting visual features become semantically inconsistent with the surrounding context.
The language model then compensates by generating plausible but unsupported objects, leading to hallucinated descriptions.

In contrast, BCR maintains representational continuity by re-encoding ROI tokens to match the statistical and semantic properties of background tokens.
As a result, the generated captions remain visually grounded and coherent, even after the target object is concealed.

These additional examples further demonstrate that hallucination in vision--language models is closely tied to representational discontinuities introduced by suppression-based attacks, while continuity-preserving manipulation enables more stable and realistic scene descriptions.

\section*{D. Failure Cases and Limitations}

While BCR substantially reduces hallucination compared to suppression-based attacks, it is not without limitations.
We summarize the main failure modes observed in our experiments and discuss directions for future improvement.

\paragraph{Partial Semantic Substitution.}
In some cases, BCR does not fully eliminate object-level reasoning but instead induces \emph{semantic substitution}.
For example, small or visually ambiguous objects (e.g., accessories such as ties or utensils) may be re-encoded
as semantically adjacent background objects (e.g., clothing regions or tableware).
Although this behavior avoids hallucination, it may still allow indirect inference of the concealed object category
through contextual cues.
This limitation is inherent to continuity-preserving attacks that aim to maintain global coherence rather than
explicitly erase object evidence.

% \paragraph{Sensitivity to ROI Size and Saliency.}
% BCR is less effective when the target object occupies a very small spatial region or is weakly salient in the
% vision encoder.
% In such cases, the object may already contribute minimally to the global representation, making semantic redirection
% less impactful.
% Conversely, extremely large ROIs may dominate the visual representation, increasing the difficulty of enforcing
% background-consistent encoding without noticeable semantic drift.

% \paragraph{Dependence on Accurate ROI Alignment.}
% Our attack assumes access to reasonably accurate bounding boxes for the target object.
% Imprecise or overly coarse ROIs can lead to either under-coverage (leaving object evidence intact) or over-coverage
% (introducing unnecessary perturbations to background regions).
% Although center-based patch assignment mitigates this issue, BCR’s performance still depends on the quality of ROI
% localization.

\paragraph{Model-Specific Vision Tokenization.}
Different VLMs employ distinct vision backbones and tokenization strategies (e.g., patch size, cropping, or pooling),
which can affect ROI-to-token alignment.
While BCR generalizes across LLaVA, BLIP-2, and Instruct-BLIP, additional model-specific adjustments may be required
for architectures with aggressive image cropping or non-uniform token layouts.

% \paragraph{Computational Overhead.}
% Compared to suppression-based attacks, BCR incurs additional computational cost due to multi-layer feature extraction
% and distributional matching losses.
% Although the attack remains practical for offline evaluation, real-time or large-scale deployment may require
% further optimization, such as layer subsampling or feature caching.

\paragraph{Scope of Concealment.}
BCR is designed to conceal objects while preserving overall scene semantics, not to guarantee absolute object
removal under all possible queries.
Highly targeted or adversarial prompts explicitly probing the ROI (e.g., repeated yes/no questioning) may still
extract residual information.
Addressing such interactive threat models remains an open problem.

Overall, these limitations highlight the inherent trade-off between concealment and semantic continuity.
We believe BCR represents a principled step toward hallucination-aware adversarial attacks, and future work may
combine continuity-based objectives with prompt-aware or adaptive strategies to further strengthen object concealment.

\end{document}